\documentclass[%
 reprint,
superscriptaddress,
 amsmath,amssymb,
 aps, prl
floatfix,
]{revtex4-2}

\usepackage{graphicx}
\usepackage[caption=false]{subfig}
\usepackage{dcolumn}
\usepackage{bm}

\usepackage{siunitx}
\usepackage{physics}
\usepackage{placeins}
\newcommand{\snr}{\text{SNR}}
\newcommand{\tem}{TEM$_{00-18}$ }

\newcommand{\Ap}{A^{\prime}}
\newcommand{\betaterm}{\frac{\beta}{\beta+1}}
\newcommand{\gagg}{g_{a\gamma \gamma}}
\newcommand{\veff}{V_{eff}}
\newcommand{\cost}{\langle \cos^2\theta \rangle_T}

\begin{document}

\preprint{APS/123-QED}

\title{Search for \SI{70}{\mu eV} Dark Photon Dark Matter with a Dielectrically-Loaded Multi-Wavelength Microwave Cavity}

\author{R. Cervantes}%
  \email[Correspondence to: ]{raphaelc@fnal.gov}
  \affiliation{University of Washington, Seattle, WA 98195, USA}
  \affiliation{currently Fermi National Accelerator Laboratory, Batavia IL 60510}
  \author{G. Carosi}
\affiliation{Lawrence Livermore National Laboratory, Livermore, CA 94550, USA}
\author{C. Hanretty}
  \affiliation{University of Washington, Seattle, WA 98195, USA}
\author{S. Kimes}
  \affiliation{University of Washington, Seattle, WA 98195, USA}
  \affiliation{currently Microsoft Quantum, Microsoft, Redmond, WA 98052, USA}
  
  \author{B. H. LaRoque}
  \affiliation{Pacific Northwest National Laboratory, Richland, WA 99354, USA}
\author{G. Leum}
  \affiliation{University of Washington, Seattle, WA 98195, USA}
\author{P. Mohapatra}
  \affiliation{University of Washington, Seattle, WA 98195, USA}
  \affiliation{currently Joby Aviation, San Carlos, CA 94063, USA}
  \author{N. S. Oblath}
  \affiliation{Pacific Northwest National Laboratory, Richland, WA 99354, USA}
\author{R. Ottens}
  \affiliation{University of Washington, Seattle, WA 98195, USA}
  \affiliation{currently NASA Goddard Space Flight Center Greenbelt, MD, United States}
\author{Y. Park}
  \affiliation{University of Washington, Seattle, WA 98195, USA}
  \affiliation{currently University of California, Berkeley, CA 94720}
  \author{G. Rybka}%
  \affiliation{University of Washington, Seattle, WA 98195, USA}
\author{J. Sinnis}
  \affiliation{University of Washington, Seattle, WA 98195, USA}
    \author{J. Yang}%
  \affiliation{University of Washington, Seattle, WA 98195, USA}
  \affiliation{currently Pacific Northwest National Laboratory, Richland, WA 99354, USA}

  \date{\today}

\begin{abstract}
  Microwave cavities have been deployed to search for bosonic dark matter candidates with masses of a few \SI{}{\mu eV}. However, the sensitivity of these cavity detectors is limited by their volume, and the traditionally-employed half-wavelength cavities suffer from a significant volume reduction at higher masses. ADMX-Orpheus mitigates this issue by operating a tunable, dielectrically-loaded cavity at a higher-order mode, which allows the detection volume to remain large. The ADMX-Orpheus inaugural run excludes dark photon dark matter with kinetic mixing angle $\chi > \num{1e-13}$ between \SI{65.5}{\mu eV} (\SI{15.8}{GHz}) and \SI{69.3}{\mu eV} (\SI{16.8}{GHz}),  marking the highest-frequency tunable microwave cavity dark matter search to date.
\end{abstract}

\maketitle


\emph{Introduction.}---There is overwhelming evidence that 84.4\% of the matter in the universe is made out of dark matter (DM)~\cite{Rubin:1982kyu, 10.1093/mnras/249.3.523,1998gravitational_lensing,10.1093/mnras/stw3385, Markevitch_2004, 2020Planck, Zyla:2020zbs}. The $\Lambda$CDM model describes dark matter as feebly interacting, non-relativistic, and stable on cosmological timescales. Not much else is known about the nature of dark matter, particularly what makes up dark matter.

The dark photon (DP) is a compelling dark matter candidate. It is a vector boson associated with an added Abelian U(1) symmetry, the simplest possible extension to the Standard Model (SM)~\cite{essig2013dark, PhysRevD.104.092016, PhysRevD.104.095029}. The dark photon, having the same quantum numbers as the SM photon, interacts with the SM photon through kinetic mixing~\cite{HOLDOM198665, HOLDOM1986196} described by the Lagrangian

\begin{align}
  \mathcal{L} = -\frac{1}{4}(F_1^{\mu \nu}F_{1\mu \nu} +F_2^{\mu \nu}F_{2\mu \nu} - 2\chi F_1^{\mu \nu}F_{2\mu \nu} - 2 m_{\Ap}^{2} A^{\prime 2}),
\end{align}
where $F_1^{\mu \nu}$ is the electromagnetic field tensor, $F_2^{\mu \nu}$ is the dark photon field tensor, $\chi$ is the kinetic mixing, $m_{\Ap}$ is the DP mass, and $\Ap$ is the DP gauge field. The photon frequency $f$ is related to the dark photon energy $E_{\Ap}$ by the relationship $f= E_{\Ap}$ (using natural units). For non-relativistic dark photons, $f\approx m_{\Ap}$.

If $\chi$ is sufficiently small, then it is stable on cosmological timescales. The lifetime is about the same as the age of the universe if $m_{A^{\prime}} (\chi^2\alpha)^{1/9} < \SI{1}{keV}$~\cite{PhysRevD.78.115012}, where $\alpha$ is the fine structure constant. This condition is easily met if $m_{\Ap} \approx \SI{e-4}{eV}$ and $\chi < \num{e-12}$.

Several mechanisms could produce cosmic dark photons, the simplest being through quantum fluctuations during inflation~\cite{PhysRevD.93.103520}. These fluctuations seed excitations in the dark photon field, resulting in the cold dark matter observed today in the form of coherent oscillations of this field. The predicted mass from this mechanism is ${m_{A^{\prime}}\approx\SI{10}{\mu eV}\left (\SI{e14}{GeV}/H_I\right )^4}$, where $H_I$ is the Hubble constant during inflation. Measurements of the cosmic microwave background tensor to scalar ratio constrain $H_I < \SI{e14}{GeV}$~\cite{2016Planck}, which makes the search for $m_{A^{\prime}}>\SI{e-5}{eV}$ well-motivated. Other mechanisms are possible and are described in~\cite{PhysRevD.104.095029, Arias_2012}.

Dark photon dark matter (DPDM) can be detected through their mixing with the SM photon. If dark photons oscillate into SM photons inside a microwave cavity with a large quality factor, then a feeble EM signal accumulates inside the cavity, which can be read by ultra-low noise electronics. This type of detector is called a haloscope and is often deployed to search for axionic DM~\cite{PhysRevLett.51.1415}. The dark photon signal power is~\cite{PhysRevD.104.092016}, in natural units, 

\begin{align}
  & P_{S} = \eta \chi^2 m_{\Ap} \rho_{\Ap} V_{eff} Q_L \betaterm L(f, f_0, Q_L)\label{eqn:dp_power}  \\
  & V_{eff} = \frac{\left (\int dV \vb{E}(\vec{x}) \vdot \vb{\Ap}(\vec{x})\right )^2}{\int dV \epsilon_r |\vb{E}(\vec{x})|^2|\vb{\Ap}(\vec{x})|^2}\label{eqn:veff}
\end{align}
 where $\eta$ is a signal attenuation factor, $\rho_{\Ap}$ is the local density of dark matter, $\veff$ is the effective volume of the cavity, $Q_L$ is the loaded quality factor, and $\beta$ is the cavity coupling coefficient. The Lorentzian term is $L(f, f_0, Q_L) = 1/(1+4\Delta^2)$, where $\Delta \equiv Q_L (f-f_0)/f_0$ is a detuning factor that depends on the SM photon frequency $f$, cavity resonant frequency $f_0$, and $Q_L$. $\veff$ is the overlap between the dark photon field $\vb{\Ap}(\vec{x})$ and the dark photon-induced electric field $\vb{E}({\vec{x}})$. Equation~\ref{eqn:dp_power} assumes the cavity size is much smaller than the DP de Broglie wavelength and the cavity bandwidth is much larger than the dark matter velocity dispersion, $Q_L << Q_{DM}$~\cite{PhysRevLett.55.1797, Kim_2020}.

The dark photon mass is unknown, so haloscopes must be tunable to search through the $\chi$ vs. $m_{\Ap}$ parameter space. The scan rate for haloscope experiments is a key figure of merit that is strongly dependent on the signal-to-noise ratio (SNR). The SNR for a haloscope signal is $\snr = (P_S/P_n)\sqrt{b \Delta t}$~\cite{doi:10.1063/1.1770483, Peng:2000hd}, where $P_n$ is the noise power, $b$ is the frequency bin width and $\Delta t$ is the integration time. $P_n$ is the combination of the cavity's blackbody radiation and the receiver's Johnson noise. The noise power can be written as $P_n=G k_b b T_n$, where $k_b$ is the Boltzmann constant, $G$ is the system gain, $T_{cav}$ is the cavity temperature, and $T_n$ is the system noise temperature referenced to the cavity. If $Q_L < Q_{DM}$, a haloscope is sensitive to dark matter within its cavity bandwidth ${\Delta f = f_0/Q_L}$. The instantaneous scan rate is then

\begin{align}
  \dv{f}{t} = \frac{\Delta f}{\Delta t} = \frac{f_0 Q_L}{b}\left (\frac{\eta \chi^2 m_{\Ap} \rho_{\Ap} \veff \beta}{\snr T_n(\beta+1)}\right )^2. 
\end{align}

Traditional haloscopes, such as those implemented by the Axion Dark Matter eXperiment (ADMX), have consisted of a right-cylindrical cavity operating at the TM$_{010}$ mode as this mode often maximizes $\veff$. ADMX currently uses this haloscope design to look for axions around a few \SI{}{\mu eV} with great success~\cite{PhysRevLett.120.151301, PhysRevLett.124.101303, PhysRevLett.127.261803}. Unfortunately, this design becomes increasingly difficult to implement at higher frequencies. Increasing mass corresponds to higher-frequency photons. Operating at the TM$_{010}$ mode would require smaller-diameter cavities. The volume scales by $V_{eff} \propto f^{-3}$ for a fixed aspect ratio, and consequently $P_S \propto f^{-3}$. This problem can be addressed by combining many cavities, as ADMX plans to do for future runs~\cite{10.1007/978-3-030-43761-9_7}. However, if the ADMX cavity's $\veff = \SI{54}{L}$ at an operating frequency $f_0 = \SI{740}{MHz}$~\cite{PhysRevLett.124.101303}, then it would be about \SI{5.4}{mL} at an operating frequency $f_0 = \SI{16}{GHz}$. Combining enough cavities to be sensitive enough to the QCD axion is challenging. This unfavorable frequency scaling motivates the design of more sophisticated resonators.

The volume can remain large if the cavity operates at a higher-order mode (as is done by the ORGAN experiment~\cite{MCALLISTER201767} to implement a \SI{26.5}{GHz} non-tunable haloscope). But higher-order modes would not couple well to dark photons since the spatial oscillations in $\vb{E}(\vec{x})$ would overlap poorly with the DP field, i.e., $\int \vb{E}(\vec{x}) \vdot \vb{\Ap} dV \approx 0$. However, dielectrics suppress electric fields and can be placed strategically to shape the electric field and increase $\veff$. With a periodic dielectric structure, the cavity can be made arbitrarily large and operate at a higher-order mode while maintaining coupling to the dark photon. This makes dielectric cavities well-suited for higher frequency dark photon searches. Because of their potential, other collaborations are developing experiments with dielectric haloscopes. Examples include MADMAX~\cite{Brun2019, PhysRevLett.118.091801}, LAMPOST~\cite{PhysRevD.98.035006, chiles2021constraints}, MuDhi~\cite{PhysRevD.105.052010}, and DBAS~\cite{PhysRevApplied.14.044051,PhysRevApplied.9.014028}.

This Letter reports results from the highest-frequency tunable microwave cavity dark matter search to date. The results exclude DPDM between \SI{65.5}{\mu eV} and \SI{69.3}{\mu eV} with kinetic mixing $\chi > \num{1e-13}$ at a 90\% confidence limit. A more detailed description of the experimental design, implementation, operation, and data analysis can be found in the companion paper~\cite{PhysRevD.106.102002}.

\begin{figure}[htp]
  \centering
  \subfloat[]{\includegraphics[width=0.45\linewidth]{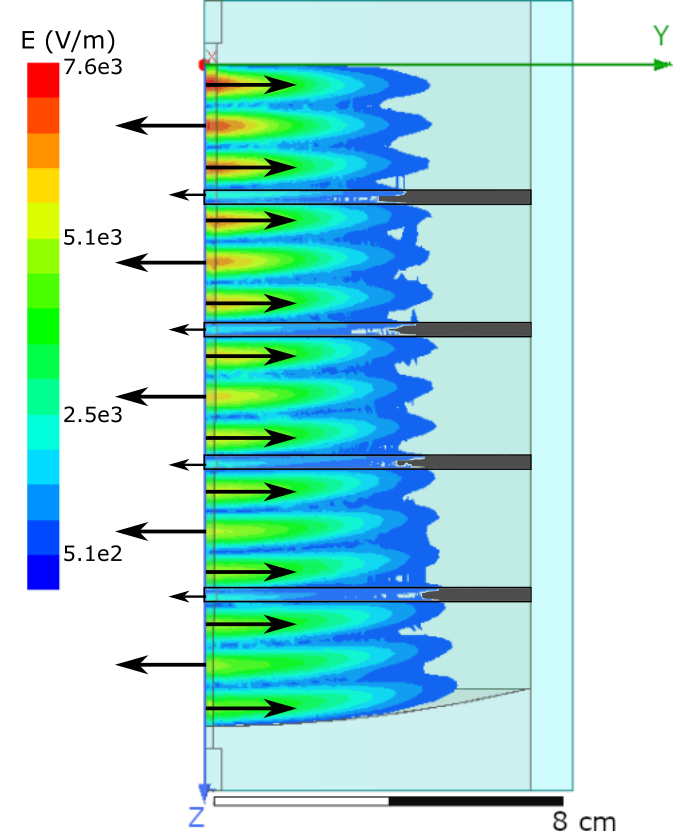}\label{fig:simulation}}\hfil
  \subfloat[]{\includegraphics[width=0.55\linewidth]{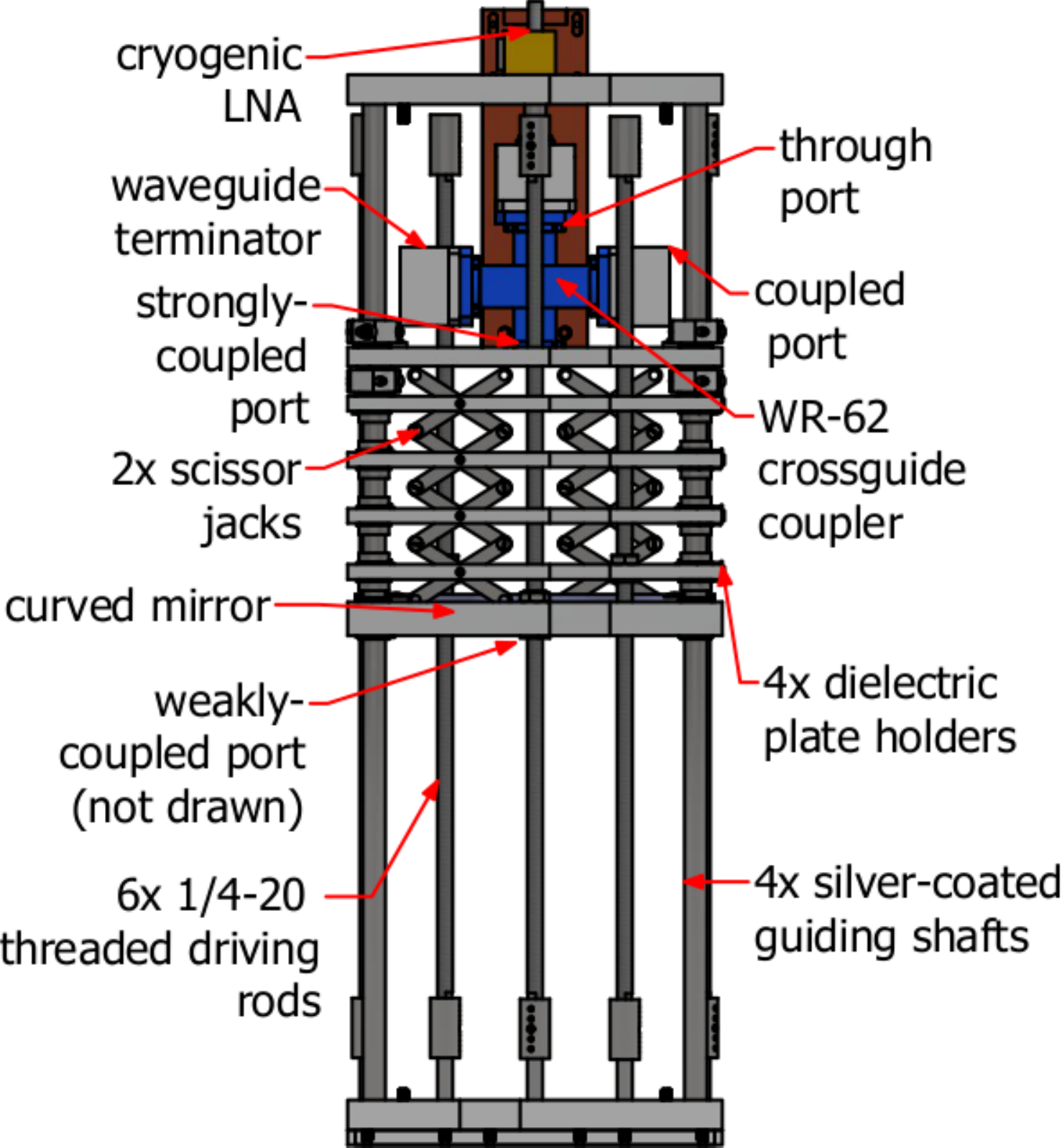}\label{fig:cad}}
  \caption{(a) The simulated electric field magnitude of \tem mode for a cavity length $L_c=\SI{15.49}{cm}$ and a corresponding resonant frequency $f_0=\SI{15.97}{GHz}$. The electric field magnitude scales linearly from blue to red in accordance with the color bar. (b) The mechanical drawings for the Orpheus cavity.}
  \label{fig:orpheus}
\end{figure}

\emph{The ADMX-Orpheus Cavity}---Orpheus implements this dielectric haloscope concept to search for dark photons around \SI{70}{\mu eV}. Orpheus~\footnote{Orpheus was initially designed to have a spatially alternating magnetic field rather than a periodic dielectric structure~\cite{PhysRevD.91.011701}. However, this alternating magnetic field design is challenging to scale to many Tesla.} is a dielectrically-loaded Fabry-Perot open cavity. The cavity operates at the \tem mode (19 antinodes across the cavity axis), and dielectrics are placed on every fourth antinode to increase the mode's coupling to the dark photon (Fig.~\ref{fig:simulation}). 

The dielectrics, purchased from Superior Technical Ceramics, consist of 99.5\% alumina sheets. Their dimensions are $\SI{15.2}{cm}\times\SI{15.2}{cm}\times\SI{3}{mm}$. The dielectric constant is $\epsilon_r = 9.8$ and the loss tangent is ${\tan\delta < 0.0001}$~\cite{stc_alumina}. A \SI{3}{mm} thickness is chosen because it is approximately half a wavelength at \SI{16.5}{GHz}.

The Fabry-Perot cavity~\cite{Clarke_1982,1447049,Dunseith_2015} consists of a flat aluminum mirror and a curved aluminum mirror with a radius of curvature, $r_0$, of \SI{33}{cm}. $r_0$ is chosen to be about twice the cavity optical length near \SI{18}{GHz} so that the flat mirror is at the focus of the curved mirror. Both mirrors are \SI{15.2}{cm} in diameter. The cavity tunes by changing the distance between mirrors, and the positions of the dielectric plates are adjusted appropriately. The curved mirror, bottom dielectric plate, and top dielectric plate are each controlled by a pair of threaded rods driven by a room-temperature stepper motor (Applied Motion Products STM23S-2EE~\cite{stm}). The scissor jacks constrain the inner two dielectric plates so that they are evenly spaced between the top and bottom dielectric plate (Fig.~\ref{fig:cad}). Thus the cavity has three degrees of freedom. 

Power is extracted from the cavity via aperture coupling connected to a WR-62 waveguide \SI{20}{dB} crossguide coupler~\cite{pasternack_crossguide}. The aperture is \SI{5.4}{mm} in diameter and \SI{3.8}{mm} thick. This was empirically determined to have an acceptable $\beta$ without too much detriment to the mechanical stability or unloaded Q ($Q_0$). $\beta\sim 1$ under cryogenic conditions. $\beta = 2$ would optimize the scan rate, but this is not attainable without making the aperture unreasonably large.

There are two sets of measurements and simulations relevant for this Letter: a room-temperature tabletop measurement that measures the cavity spectrum and the cryogenically-cooled DPDM haloscope search. These measurements differ in two major aspects. First, the dielectric dissipation is substantially reduced in the cryogenic measurement, which increases $Q_L$ and $\beta$. Second, the room-temperature measurements followed a configuration where the dielectric plates were evenly-spaced throughout the cavity for each tuning step. However, in the cryogenic search, the dielectric plate positions deviated from the evenly-spaced configuration. This deviation was caused by motor tuning issues and misconfigured software. The \tem mode was resimulated with the dielectric positions measured in the dark photon search. These two sets of measurements and simulations are hereafter referred to as either room-temperature or cryogenic.

\begin{figure}[htp]
  \centering
  \includegraphics[width=0.45\textwidth]{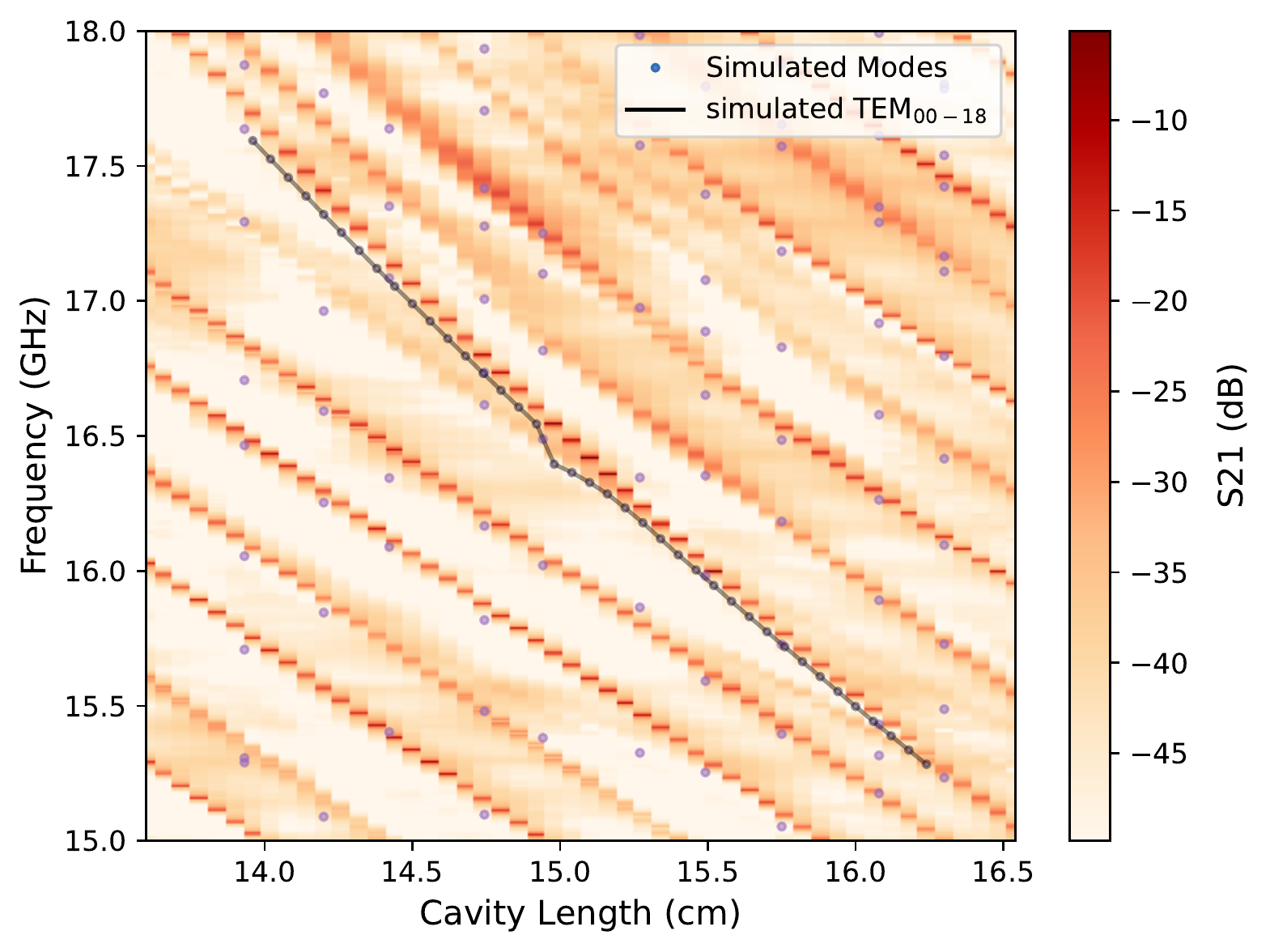}
  \caption{Orpheus mode map with the simulated \tem mode overlaid. Both measurement and simulation correspond to a room-temperature tabletop setup in which the dielectrics maintained even spacing throughout the cavity. This configuration suffers from a mode crossing at about \SI{16.4}{GHz}. This mode crossing was mitigated in the dark photon search by deviating from the evenly-spaced configuration.}
  \label{fig:orpheus_modemap}
\end{figure}

The cavity spectrum is measured with the room-temperature setup and is visualized with a mode map, a 2D plot of the transmitted power through the cavity as a function of frequency and cavity length, as shown in Fig.~\ref{fig:orpheus_modemap}. The dark lines are the resonant modes. The room-temperature simulation of the \tem mode is overlaid on the mode map and is found to agree with the measured mode often to better than one part-per-thousand within the cryogenic search tuning range. The exception is at the mode crossing around \SI{16.4}{GHz}. The simulation predicts a kink in the mode frequency at the mode crossing, but the measured mode is shown to tune smoothly and continuously. This suggests that the simulation overestimates the effect of the mode crossing, possibly because the mode crossing requires more resolution to simulate accurately. Regardless, the mode crossing was mitigated in the cryogenic dark photon search because the dielectric plates deviated from the evenly-spaced configuration. 

Orpheus's sensitivity to the dark photon is determined from the cavity's $f_0$, $\veff$, $Q_L$, $\beta$ (Equation~\ref{eqn:dp_power}). The crux of the Orpheus experiment is to increase $\veff$ using the dielectric structure. Since $\vb{E}$ cannot be measured directly, $\vb{E}$ is simulated using Finite Element Analysis simulation software (specifically, ANSYS\textsuperscript{\textregistered} HFSS 2021 R1). The field is simulated for the cryogenic search and used to calculate $\veff$ (Equation~\ref{eqn:veff}). Because of the orientation of the WR-62 waveguide, the receiver is only sensitive to $\vb{E}_y$ (one of the transverse coordinates), so ${\veff = \left (\int dV \vb{E}_y(\vec{x})\right)^2/(\int dV \epsilon_r |\vb{E}_y(\vec{x})|^2)\langle \cos^2\theta \rangle_T}$, where $\theta$ is the angle between  the electric field along $\hat{y}$ and the dark photon field. $\theta$ is unknown, but $\cost =1/3$ if the dark photon is randomly polarized~\cite{Arias_2012, PhysRevD.104.092016, PhysRevD.104.095029}. 

\begin{figure}[htp]
  \includegraphics[width=0.45\textwidth]{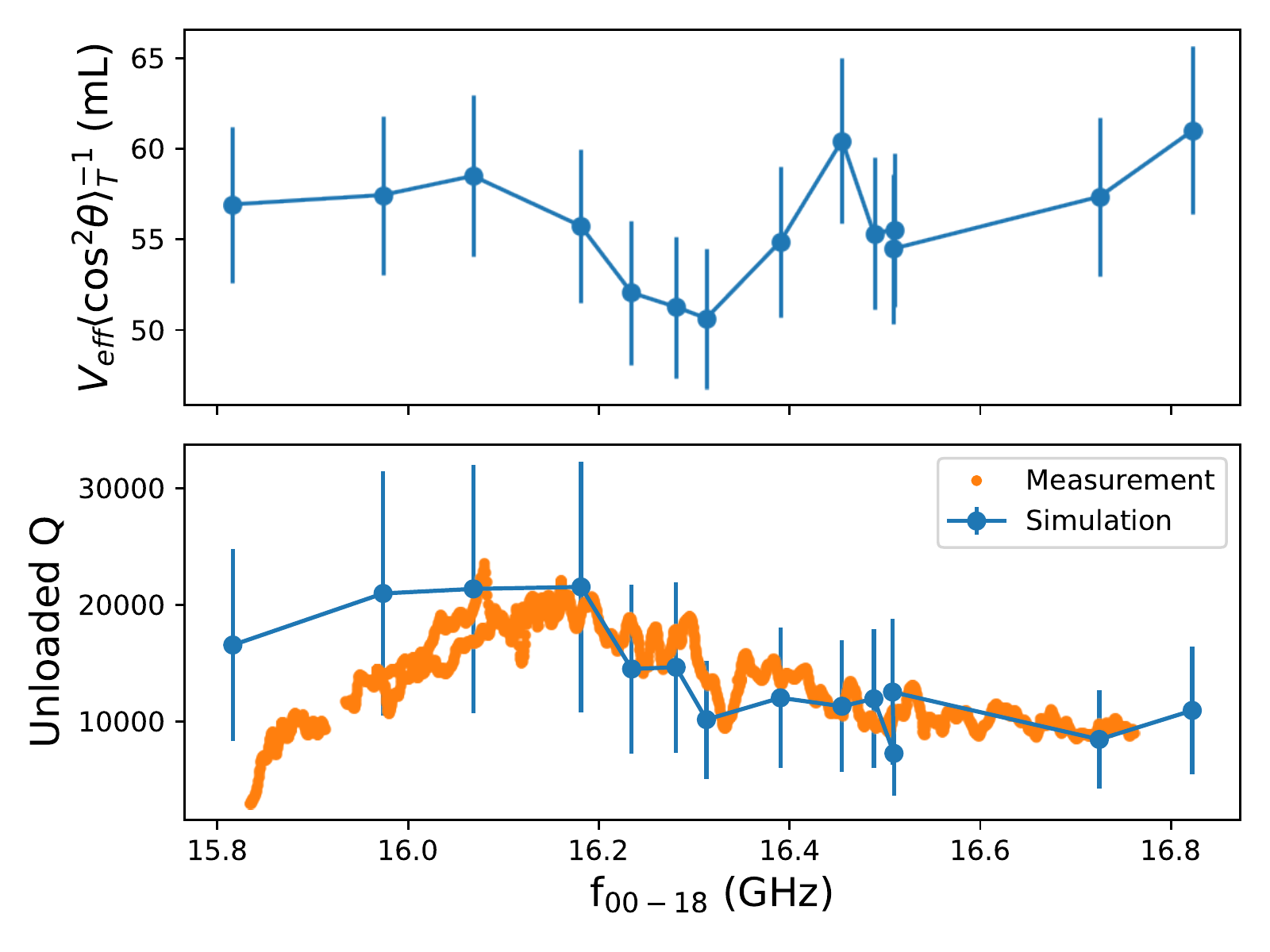}\label{fig:mode_param}
  \caption{$\veff$ and $Q_0$ as a function of mode frequency. The simulated $Q_0$ is consistent with the measured $Q_0$ within the simulated uncertainties. The measured $Q$ is double-valued at some frequencies because of the hysteresis in the tuning mechanism.}
  \label{fig:orpheus_characterization}
\end{figure}

The simulated field is shown in Fig.~\ref{fig:simulation}, and cryogenic simulation of $\veff$ and $Q_0$ is shown in Fig.~\ref{fig:orpheus_characterization}. $\veff\cost^{-1}\sim\SI{55}{mL}$ for much of the tuning range, which is about a factor of 10 times larger than the ADMX cavity rescaled to operate the same frequency. After the dark photon search concluded, it was discovered that deviations from the evenly-spaced configuration serendipitously increased $\veff$ and mitigated a problematic mode crossing (more detail in~\cite{OrpheusPRD}). The relative uncertainty in $\veff$ is 7.14\%. This uncertainty is determined by simulating how $\veff$ is affected by possible misalignments of the mirrors and dielectric plates, uncertainty in dielectric constant and loss tangent, and the effects of the mechanical structure~\cite{OrpheusPRD}. Simulating these perturbations also caused the simulated $Q_0$ to vary by 50\%. Fig.~\ref{fig:orpheus_characterization} shows that within the uncertainty of the cryogenic simulation, $Q_0$ matches the measured $Q_0$ determined from the measured $Q_L$ and $\beta$, $Q_0 = Q_L (1 + \beta)$. This matching $Q_0$ corroborates the simulated $\veff$.

The measured $Q_0$ drops off below \SI{16}{GHz} and above \SI{16.2}{GHz}, suggesting Orpheus has a natural bandwidth. This is because the fixed dielectric thickness and mirror radius of curvature are optimal for a small range of frequencies. These parameters can be adjusted to allow Orpheus to scan for dark matter at different frequencies.

\begin{figure}
  \centering
  \subfloat{\includegraphics[width=\linewidth]{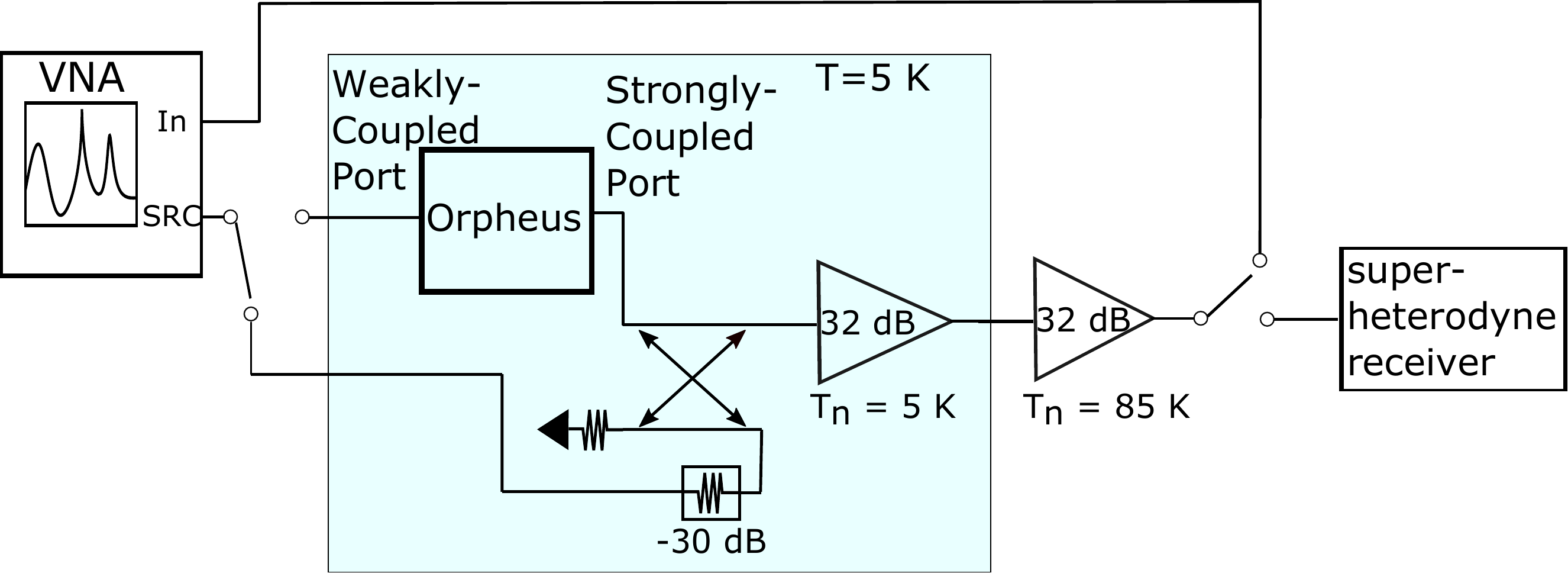}\label{fig:cold_electronics}}\\
  \subfloat{\includegraphics[width=\linewidth]{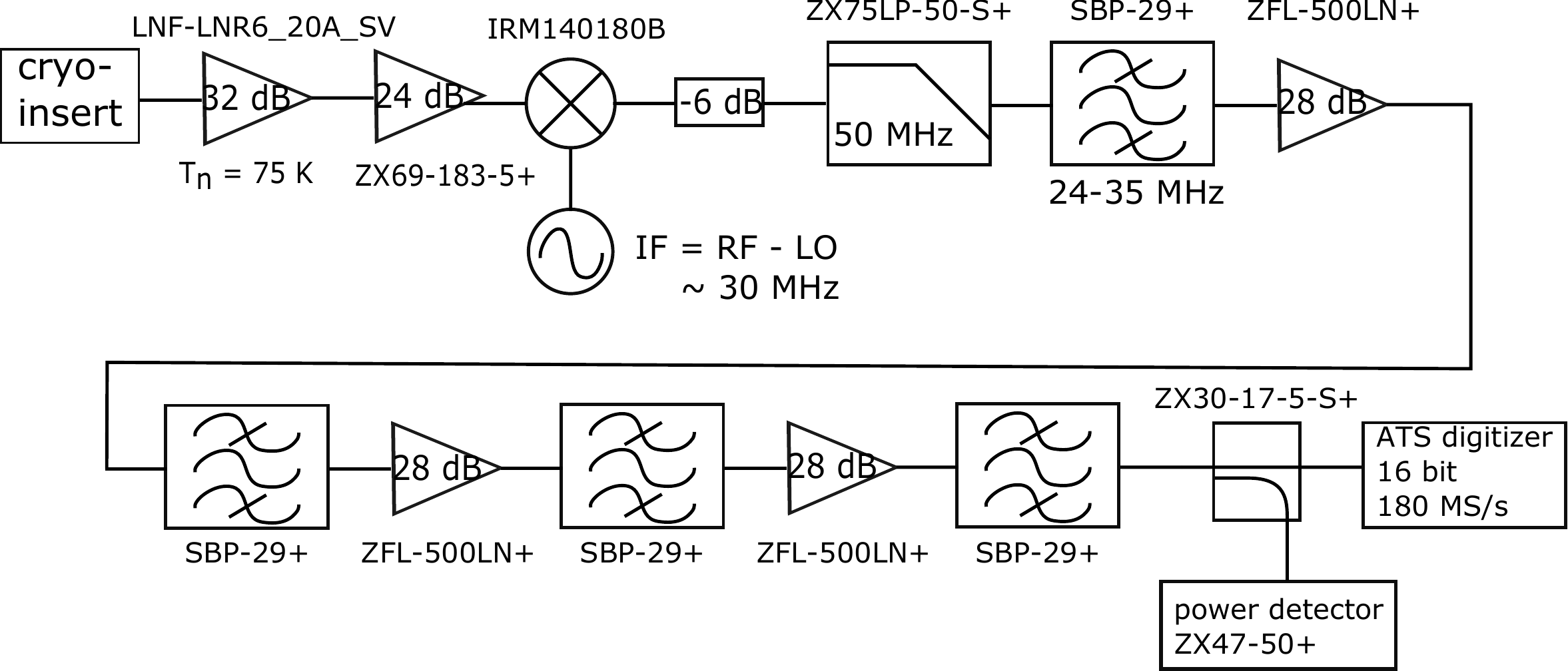}}
  \caption{Top: A diagram of how transmission, reflection, and power measurements are taken. Room temperature Teledyne switches are used to switch between transmission and reflection measurements, and from vector network analyzer (VNA) measurements to power measurements. The directional coupler allows the VNA source to bypass the amplifier for a reflection measurement. Bottom: Diagram of the superheterodyne receiver.}
  \label{fig:electronics}
\end{figure}

\emph{Dark Photon Search Experimental Setup}---The cavity is cooled down to liquid helium temperatures. The power of the cavity is first amplified by a cryogenic heterostructure
field effect transistor (HFET) amplifier (LNF-LNC6 20C~\cite{lnf}) and then is processed by the superheterodyne receiver in Fig.~\ref{fig:electronics}.

The search strategy is to tune the cavity to scan for dark photons with different $m_{\Ap}$. For each cavity length, a series of ancillary measurements are taken to extract a noise power calibration and expected dark photon signal power. These measurements include the cavity length, dielectric positions, cavity temperature, transmission coefficient, and reflection coefficient. The cavity temperature is used to determine the noise power, and the transmission and reflection coefficients are used to extract $f_0$, $Q_L$, and $\beta$. The power spectrum is then measured out of the cavity for either \SI{30}{s} or \SI{100}{s}, depending on the desired SNR. The dark matter signal would be observed in the power spectra as a spectrally-narrow power excess over the noise floor.

\emph{Analysis and Results}---For a critically-coupled Orpheus cavity operating in the Rayleigh-Jeans limit ($k_bT_{cav} >> hf$), the system noise temperature is modeled as ${T_n = T_{cav} + T_{rec}},$
where $T_{cav}$ is the physical temperature of the cavity, and $T_{rec}$ is the noise temperature of the receiver chain from the output of the cryogenic amplifier outward (see~\cite{OrpheusPRD} for more details). $T_{cav}$ is measured using a pair of calibrated Cernox resistors and is typically \SI{4.7 \pm 0.1}{K}. $T_{rec}$ is dominated by output noise temperature of the 1st stage amplifier $T_{amp}$, and is more accurately obtained by the Friis cascade equation~\cite{friis} (future runs will incorporate an in-situ measurement of $T_{rec}$). From the manufacturer's calibration~\cite{lnf}, $T_{rec} = \SI{5.0 \pm 0.5}{K}$. This results in $T_n \sim \SI{9.7}{K}$. 

The cavity length and position of the dielectrics were calculated using the motor encoder values. However, there is a systematic offset between measured and simulated resonant frequency for a given cavity length. This frequency offset is removed by adding \SI{0.7}{mm} to the measured cavity length derived from the motor encoder values.  This systematic bias may be caused by mechanical contractions during cooldown or by tuning hysteresis. After accounting for the systematic bias, the measured $f_0$ matches the simulated $f_0$ often by less than one part per thousand.

The data collected between 9/3/2021 and 9/7/2021 are used to search for dark photons between \SI{65.5}{\mu eV} (\SI{15.8}{GHz}) and \SI{69.5}{\mu eV} (\SI{16.8}{GHz}). The system noise temperature $T_n$ is used to calibrate the power excess. All measured power excesses are consistent with Gaussian noise, so a 90\% confidence level exclusion limit is placed on the kinetic mixing $\chi$ in this mass range. The procedure for deriving the exclusion limits follows the procedure developed by ADMX and HAYSTAC~\cite{PhysRevD.64.092003, PhysRevD.96.123008, PhysRevD.103.032002}, and is adapted for dark photon searches~\cite{PhysRevD.104.095029, PhysRevD.104.092016, OrpheusPRD, cervantes2021search}. The analysis for this measurement is described in more detail in the companion paper~\cite{OrpheusPRD}.

\begin{figure}
  \centering
  \subfloat{\includegraphics[width=\linewidth]{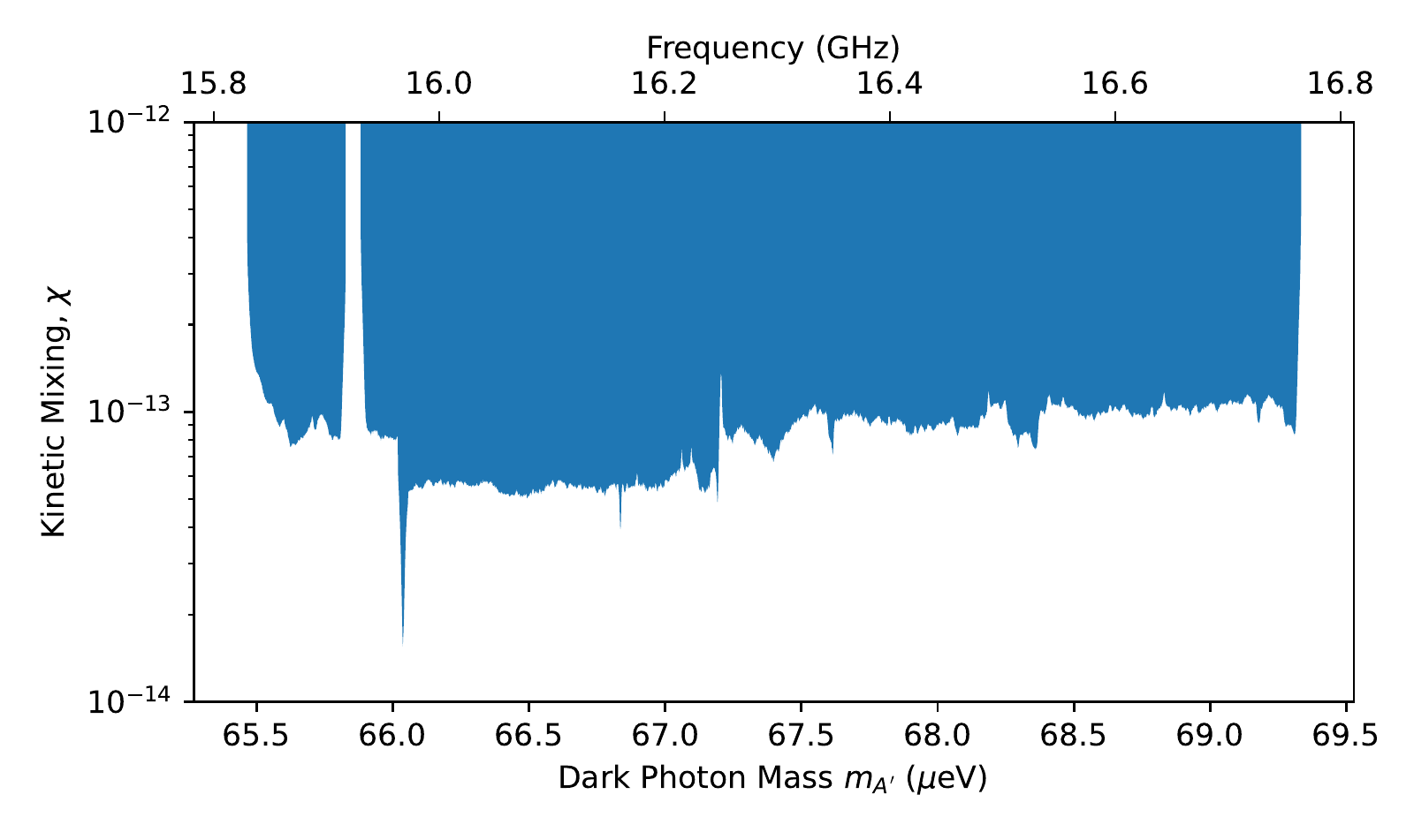}}\\
  \subfloat{\includegraphics[width=\linewidth]{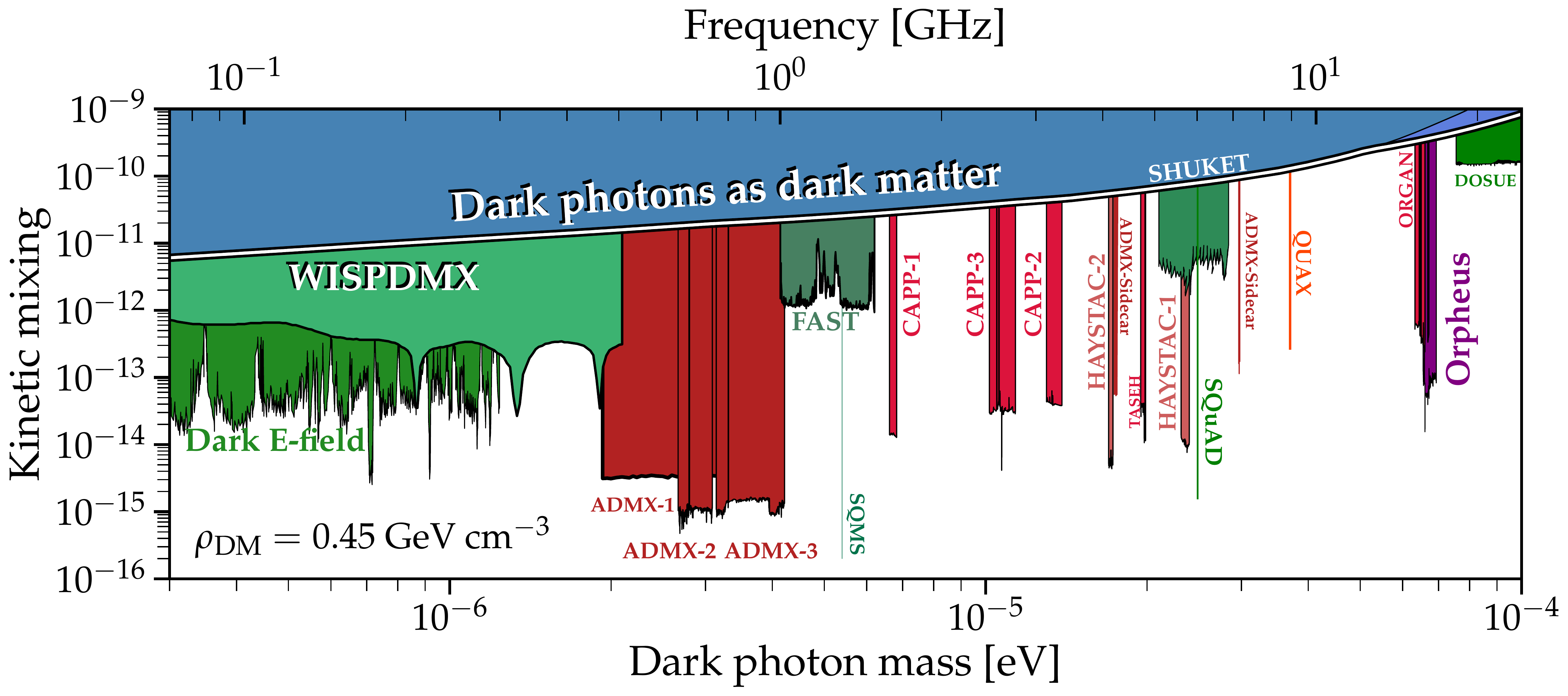}}
  \caption{Top: A 90\% exclusion on the kinetic mixing parameter space. Bottom: Orpheus limits in the context of other microwave cavity haloscopes. Figure adapted from~\cite{ciaran_o_hare_2020_3932430}.}
  \label{fig:dp_limits}
\end{figure}

The derived 90\% exclusion of dark photons is plotted in Fig.~\ref{fig:dp_limits}, assuming dark photons are randomly polarized $\cost = 1/3$. If dark photons are polarized across cosmological horizons, the scenario implies $\langle \cos^2\theta \rangle \geq 0.076$ for a 90\% confidence limit, and the results can be appropriately rescaled~\cite{PhysRevD.104.095029, PhysRevD.104.092016}.

\emph{Conclusion and Discussion}---Orpheus has excluded DPDM higher in frequency than other haloscope experiments while also having a respectable tuning range. Orpheus also demonstrates the potential advantages over a cylindrical haloscope operating at similar frequencies such as ORGAN~\cite{doi:10.1126/sciadv.abq3765}. Orpheus has three times $\veff$ and $Q_L$ compared to ORGAN, and achieved as much as almost an order of magnitude better in sensitivity with less experimental run time (Fig.~\ref{fig:dp_limits}). A \SI{1.5}{T} dipole magnet is currently being fabricated that would allow Orpheus to search for axions as well as DPDM. Applying a hypothetical \SI{1.5}{T} magnet to the dark photon data suggests Orpheus would be sensitive to axions with $\gagg \sim \SI{3e-12}{GeV^{-1}}$ from \SI{15.8}{GHz} to \SI{16.8}{GHz}. With more experimental iterations with different dielectric thicknesses and mirror radius of curvatures, Orpheus can potentially scan the axion and dark photon parameter space higher than $\sim \SI{10}{GHz}$. 

Orpheus lays the groundwork for other future dielectric array experiments such as MADMAX. It demonstrates the feasibility and tolerance of the tuning mechanism. Orpheus can also become sensitive to the QCD axion by making it larger and colder. With the same integration time, Orpheus can achieve Kim-Shifman-Vainshtein-Zakharov (KSVZ) sensitivity if ${V_{eff}\sim \SI{120}{mL}}$ and $Q_L\sim \num{2e4}$, $T_n\sim\SI{1}{K}$, and $B_0 = \SI{10}{T}$. That would require the electromagnetic optimizations that increase $\veff$ and reduce diffraction losses, cooling the cavity with a dilution refrigerator, quantum noise limited amplifiers, and technological advances in winding superconducting dipole magnets. Dine-Fischler-Srednicki-Zhitnitsk (DFSZ) sensitivity may be reached by increasing the cavity size to ${V_{eff}\sim \SI{600}{mL}}$. Detection mechanisms that subvert the Standard Quantum Limit, such as vacuum squeezing~\cite{Backes2021} and superconducting qubit photon counters~\cite{PhysRevLett.126.141302} would be advantageous in this frequency range for increasing sensitivity.

\emph{Acknowledgements}---This work was supported by the U.S. Department of Energy through Grants No. DE-SC0011665 and by the Heising-Simons Foundation. Pacific Northwest National Laboratory (PNNL) is operated by Battelle Memorial Institute for the DOE under Contract No. DE-AC05-76RL01830. Prepared by LLNL under Contract DE-AC52-07NA27344 with release \#: LLNL-JRNL-834494. Many parts were fabricated by the University of Washington Physics Machine Shop and CENPA machine shop. CENPA administration and engineers helped develop the infrastructure to commission the Orpheus test stand. Finally, we thank M. Baryakhtar for helpful discussions and clarification on dark photon cosmology.

\bibliography{orpheus_thesis}

\end{document}